\documentclass[aps,preprint,     showpacs,groupedaddress]{revtex4}

\usepackage{graphicx}         
\usepackage{bm}               
\usepackage{amssymb}          
\usepackage{amsmath}          

\begin{document}

\title{Why is the Dark Axion Mass $10^{-22}$ eV?}


\author{Tzihong Chiueh}
\email{chiuehth@phys.ntu.edu.tw}
\affiliation{Department of Physics, National Taiwan University, Taipei 10617, Taiwan}
\affiliation{National Center for Theoretical Sciences, National Taiwan University, Taipei 10617, Taiwan}


\begin{abstract}
Scalar field dark matter likely is able to solve the small-scale cosmology problems facing the cold dark matter (CDM), and has become an emerging contender to challenge the CDM.   It however requires a particle mass $\sim 1 - 2 \times10^{-22}$eV.  We find such an extremely small particle mass can naturally arise from a non-QCD axion mechanism,
under fairly general assumptions that the axion is the dominant dark matter, and a few species of self-interacting light particles of comparable masses and a massless gauge boson both decouple from the bright sector since the photon temperature exceeds 200 GeV.  These assumptions also set the axion decay constant scale to several $\times 10^{16}$ GeV.  Given the above axion mass, we further pin down the dark-sector particles to consist of single-handed fermion and anti-fermion.  With a mass around $92-128$ eV, these dark-sector fermions may constitute a minority population of dark matter.  In the simplest SU(2) gauge field model, a dilute instanton gas as dark photons can contribute to about $2.5\%$ of the total relativistic relics in the cosmic microwave background radiation.

\end{abstract}

\pacs{14.80.Va, 98.80.Cq, 98.35Gi}
\maketitle

\section{Introduction}
\label{sec:1}
The scalar field dark matter has been in recent years gaining supports as a viable dark matter candidate \cite {hu,woo,sch1,sch2,lee,lor,har,mag,rob,mar}. In particular, the complex scalar field, dubbed $\psi$DM, of boson mass $m_a= 10^{-22}$eV has been shown to solve the most pressing small-scale cosmology problems, i.e., the core-cusp problem and the missing satellite problem of CDM \cite{sch1}.  It also solves the catch-22 problem facing the warm dark matter \cite{mac}. These are all small-scale problems associated with small galaxies.  Observations of Lyman-$\alpha$ forests place another constraint on the next scale about $1$Mpc (co-moving coordinate) for alternative dark matter models \cite{sel, vie}. $\psi$DM of $m_a\approx 10^{-22}$eV has been shown to be marginally consistent with the data of Lyman-$\alpha$ forests \cite{mar}, but an increase of particle mass by a factor 2 will definitely ease the tension.  
 
From astrophysics standpoint, $\psi$DM offers a new window of opportunity for quantitative comparisons with more unsolved problems.  But from particle physics standpoint, it remains enigmatic for a particle that is so light.  String theories have predicted an axiverse with a hierarchy of light particle masses \cite{asi,ach}.  The axiverse has so far 
gained no support from astronomical observations, which
tend to favor a narrow range of mass scale.  Despite this, the axion mechanism is still very attractive, in that it has almost no tunable parameter in some regime, thus making the predictions solid and definitive.  

Axion mechanism is originally proposed by Peccei and Quinn to solve the CP problem in strong interactions \cite{pec1,pec2}. 
When the axion decay constant $f_a$ is taken to be around 100 GeV for the Higgs field in the original proposals, axions of $0.1-1$ MeV mass were predicted \cite{wei, wil}, and no such axions were detected in experiments.  The Higgs field is later replaced by an exotic large field, $f_a=10^{11-13}$ GeV, and due to the large field suppression, axions become so weakly interacting that they are invisible \cite{kim,shi,din,zhi}.  The invisible axion is a viable CDM candidate.

In both Peccei-Quinn and invisible axion models, axions acquire mass from a coupling to the QCD gauge boson and $\sqrt{m_af_a} \sim 100$ MeV \cite{kim1}.  Here we take a different path and consider a non-QCD axion.   In our model, we only assume that a few species of light self-interacting particles in the dark sector, which
help generate axions, get decoupled from Standard Model (SM) particles ever since $T =200$ GeV, and that the axion is the primary dark matter in the present universe, although
the relics of these light particles may survive annihilation as a minority dark matter component.  These two assumptions set the desired axion mass scale and the axion decay constant scale. 

The Peccei-Quinn mechanism involves a tilt of the originally axisymmetric potential of a very large field $f_a$.  The potential tilt is associated with the instanton-anti-instanton imbalance, where the instanton interaction Lagrangian becomes $\theta F^{\mu\nu}_i \bar F_{\mu\nu i}\propto \delta(1-\cos(\theta))$,  acting as the axion potential. Here $\delta=\Lambda^4/f_a^4$ is the tilt angle and $\Lambda=(m_a f_a)^{1/2}$.
The tilt angle brings about another energy scale $\Lambda$, and it can be the mass scale of these light dark particles.

The finite-temperature axion potential can assume the form $m^2(T)(1-\cos(\theta))$\cite{bae}, where $m^2(T)$ represents the smooth thermal turn on of the potential.
Define $T_m$ to be the temperature when $3H=m_a$.   
Solutions in two different parameter regimes, $\Lambda \geq T_m$ (cold-start regime) and the opposite (hot-start regime), oscillate very differently.   But when $\Lambda/T(a) \gg 1$, both solutions asymptotically approach $a^{-3/2}$, and thus the axion with energy density decreaseing as $a^{-3}$ mimics a CDM.  

All solutions of the cold-start regime are almost identical and oscillations begin at $3H(a)=m_a$.  Solutions of the hot-start regime begin oscillations also when $3H(a)=m(T)$, but when $T_m \gg \Lambda$ the oscillation can start very early on when $m(T) \ll m_a$.  The "hot" solution obeys $m(T)\langle\theta^2\rangle (a/a_m)^3 \approx \beta$, an adiabatic invariant arising from the conservation of wave action $\langle \theta d\theta/dt\rangle a^3$. Here $\beta$ is a constant of time but its value varies depending on the details of $m(T)$.  
The QCD axion is in this hot-start regime.   Fortunately, our case is in the much simpler cold-start regime since the axion in question is very light.  The simplicity of cold-start permits this axion model to yield rather definitive predictions.

\section{Axion mass scale}
\label{sec:2}

To obtain the axion mass scale and the energy scale of the axion decay constant,
we first work backward in time in the radiation dominant era, starting from the radiation-matter equality where the axion energy density $(1/2)m_a^2f_a^2 \theta^2$ equals the radiation energy density at $T_{eq}\sim 0.73$ eV, or $\Omega_m=0.3$.   
  At the onset of mass oscillation, we have 
\begin{equation}
(6/5)(1/2)m_a^2 f_a^2\theta_0^2 = q  (T_m/T_{eq})^3\rho_{eq},
\end{equation}
where $\theta_0$ is the initial value of the axion field and $q$ is the axion fraction of the dark matter, $100\%$ for $q=1$.
Here $6/5$ accounts for the $1/6$ baryon contribution to the cold matter, $\theta_0$ is the initial axion field amplitude and $\rho_{eq}=(g_{eq}\pi^2/30) T_{eq}^4 $, the radiation (photon+neutrino) energy density at radiation-matter equality with the degree of freedom $g_{eq}=3.36$.  

On the other hand, $H_m/H_{eq} = m_a/3H_{eq}$ and $H_m/H_{eq}=(T_m/T_{eq})^2$.  Thus it follows
\begin{equation}
T_m =T_{eq}(m_a/3H_{eq})^{1/2}
\end{equation}
Substituting $T_m$ in Eq.(2) into Eq.(1), we find
\begin{equation}
\begin{split}
m_a f_a^4& = (1.84/q)^2 (3H_{eq})^{-3} (\theta_0 ^{-1}T_{eq}^2)^4  \\
&=2.7\times 10^{81} h^{-3} q^{-2}(\theta_0^{-1/2}T_{eq})^8 eV^{-3}, 
\end{split}
\end{equation}
where $H_{eq} =3.6\times 10^{-28} h$ eV and $h=H_0/100$km/s/Mpc have been used to obtain the second equality.
The factor $1.84=2(5/6)(3.36\pi^2/30)$ comes from Eq.(1).  Equation (3) is a model-independent result in the cold-start regime. 

Next, we proceed forward in time starting from high energy before the axion oscillation, and we pursue a specific axion creation scenario in the dark sector. The partition of energy density between the bright sector and the dark sector is proportional to the degrees of freedom.  The bright sector contains all SM particles at $T=200$ GeV and has the effective degrees of freedom $g=106.8$.  On the other hand, we assume that the dark sector light particles already get decoupled from the bright sector and have $g_d$ degrees of freedom at $T= 200$ GeV.  Due to particle annihilation, the bright sector eventually contains 2 photon degrees of freedom below $500$keV.   By contrast, during 
the time between $T=200$ GeV and $T = 500$ keV, the dark sector has not had particle annihilation, since these dark light particles are still relativistic. 

Conservation of entropy demands that the bright sector photon temperature be raised by a factor $(106.8/2)^{1/3}$ compared to the dark sector temperature.  But this needs to be corrected for neutrino decoupling at $\sim$ MeV, where
the original $g = 2+(7/8)(4+6)$ becomes $g=2+(7/8)(4)$.  This brings down the temperature ratio to 
$(27.3)^{1/3}$.  We let $\bar T$ be the dark sector temperature, and $\bar T = (1/27.3)^{1/3} T$ for $T<500$ keV.

The axion mass
oscillation occurs at $T_m$, and
we now match the two conditions derived from backward in time and from forward in time at $T=T_m$.  
The ratio of energy densities between the dark sector and the bright sector at $T_m$ is $\rho_{d,m}/\rho_m = (g_d/g_{eq})(1/27.3)^{4/3}$.  At the onset of the axion oscillation, a fraction $r$ of the dark sector particle energy must be converted to create the axion potential.  Thus the axion initial potential energy density at $T_m$ is
\begin{equation}
(1/2)m_a^2 f_a^2 \theta_0^{2}=(1/27.3)^{4/3}(rg_d)(\pi^2/30)T_m^4.
\end{equation}

Inverting $T_m$ from Eq. (1) and substituting into Eq.(4), one finds $\Lambda(\equiv(m_af_a)^{1/2})$ becomes
\begin{equation}
\begin{split}
\Lambda&= (27.3)(1.84/q)(2\pi^2/30)^{-3/4} (rg_d)^{-3/4}(\theta_0^{-1/2} T_{eq}) \\
&=68(rg_d)^{-3/4}q^{-1}(\theta_0^{-1/2} T_{eq}),
\end{split}
\end{equation}
Equations (2.3) and (2.5) can combine to yield
\begin{equation}
\begin{split}
m_a &= (27.3)^{8/3}(1.84/q)^{2}(2\pi^2/30)^{-2} ( 3H_{eq}) (rg_d)^{-2} \\
&= 0.56 \times10^{-22} h (rg_d)^{-2} q^{-2} eV
\end{split}
\end{equation}
and
\begin{equation}
\begin{split}
f_a &=(27.3)^{-2/3} (2\pi^2/30)^{1/2}(3H_{eq})^{-1}(rg_d)^{1/2}\theta_0^{-1} T_{eq}^2  \\
&=8.4\times 10^{16}(rg_d)^{1/2} h^{-1}(\theta_0^{-1/2}T_{eq}/eV)^2 GeV.
\end{split}
\end{equation}  
From Eq.(2), we can also obtain the temperature in the bright sector 
\begin{equation}
\begin{split}
T_m&= (27.3)^{4/3}(1.84/q)(2\pi^2/30)^{-1} (rg_d)^{-1} T_{eq}
\\
&=225 (r g_d)^{-1} q^{-1}T_{eq},
\end{split}
\end{equation} 
and that in the dark sector 
\begin{equation} 
\bar T_m=(27.3)^{-1/3} T_m= 75 (rg_d)^{-1} q^{-1}T_{eq}.  
\end{equation}

Equations (5) - (9) mark our main results, which depend on one relevant factor $r g_d$.  
In the next section we shall argue that $r g_d$ cannot be far from unity if axions are to be 
the primary dark matter.   

Before doing that, we shall examine other energies given the observed axion mass constraints.
The standard values for $h=0.67$ and $T_{eq}=0.73$eV are useful to evaluate the above numbers.  
We assume the dark matter is entirely composed of axions, thus $q=1$.  
When the factor $rg_d=0.6-0.43$, we recover the $\psi$DM boson mass $\sim 1 - 2\times10^{-22}$eV determined from dwarf galaxy observations and constrained by Lyman-$\alpha$ forest data.  Comparing Eqs. (5) and (9), $\bar T_m$ is comparable to $\Lambda$ for a wide range of $rg_d$, indicating that the two are largely locked together.  For $rg_d =0.6 - 0.43$ and $\theta_0\sim 1$, we find that $\Lambda=71-94$ eV and $\bar T_m=92-128$ eV, and 
the axion decay constant $f_a=3.9-4.6\times 10^{16}$ GeV, not far above the GUT scale.

What do $\theta_0$ and $r g_d$ mean?
The fudge factor $\theta_0 \sim 1$ is of anthropic origin, reflecting the fact that we happen to live in a patch of the universe with $\Omega_m h^2=0.13$ thus $T_{eq}\sim 0.73$ eV, and in a different patch of the universe where $\theta_0$ is different, $T_{eq}$ will be different to keep the same physics, i.e., $m_a$ and $\Lambda$.
Hence $\theta_0$ has no physical relevance.
However, the factor $r g_d$ has particle physics significance.
This factor can be interpreted as the degrees of freedom given to the axion potential energy density 
measured in unit dark radiation energy density.  

\section{Dark axion model}
\label{sec:3}
Let us suppose the dark sector to consist of weakly-interacting fermions with a mixture of Dirac and Majorana masses so that they contain a finite CP violating phase $\phi$, and the massless $SU(2)$ gauge bosons that form instantons and gauge couple to fermions.  The instanton ground state is a superposition of a series of individual instantons of different positive and negative integer winding numbers, i.e., a dilute instanton gas.
When fermions annihilate, the CP violating phase $\phi$ must go somewhere.  The instanton gas would need to adjust to generate an energetically unfavorable fractional averaged winding number to accommodate the arbitrary CP violating $\phi$.  To ease the tension, a third dynamical field $\bar\theta$ therefore arises and in place of instantons to absorb $\phi$.  With the presence of $\bar\theta$, the instantons have an average winding number $\propto\bar\theta+\phi$, which is energetically favorable since $\langle (\bar\theta+\phi)\rangle = 0$.  Here, $\bar\theta+\phi(\equiv \theta)$ is the axion field and proportional to the force acting on the axion; hence the axion potential is $\propto \theta^2$.  

When one further identifies $\theta$ to be the angle of the large field, one immediately realizes that it is the CP violating phase $\phi$ of dark-sector fermions that breaks the rotational symmetry and induces a tilt to the originally axisymmetric potential of the large field.  Thus, the action of transferring $\phi$ is in between the dark-sector fermions and the axions;  the instanton gas is merely a background medium that is passively perturbed by the release of $\phi$.  
Hence when CP violating fermions annihilate, the released rest mass energy must largely be converted to the axion potential to produce resonance oscillation, and leaves little to share with the instanton random excitations as heat.\footnote{Although the underlying axion scenario is similar to QCD, the present model is in great contrast to QCD. Instantons in QCD are tightly coupled to quarks, and when the latter are annihilated the former will no longer exist.  But in the present model with weak interactions, an almost free instanton gas is still present as dark photons even in the absence of dark fermions.}  

The above picture is in the zero temperature limit.  
At finite temperature, the
picture does not change, except one must distinguish the rest mass energy that is converted to the axion potential and
the kinetic energy that is dumped to the heat bath.   The rest mass energy is released at the time when particles become trans-relativistic, and we have indeed seen $\Lambda \sim \bar T_m$ for a wide range of $r g_d$, which is strongly suggestive that dark particles have masses $\sim \Lambda$ and they are just becoming trans-relativistic to make the energy conversion.  

To calculate the rest mass (or binding) energy density released during the annihilation, we
can again make use of entropy conservation.   Consider an amount of matter and antimatter annihilated in a static box.  The excess energy density released by annihilation is proportional to $g_p((g_p/g_q)^{1/3}-1)$, where $g_p$ is the initial degree of freedom that decreases to $g_q$.  Hence the fractional excess energy density per annihilation is
\begin{equation}
({30\over {\pi^2 \bar T^4}}){\Delta\rho_d \over \Delta g_d}=g_{d,p}{{(g_{d,p} /g_{d,q})^{1/3}-1}\over {g_{d,p}-g_{d,q}}}\to{1\over 3},
\end{equation}
as $q\to p$.
This $1/3$ fraction originally yields a rise of thermal energy due to the release of rest mass energy, but it now needs to work on the creation of axions in such a way that the thermal energy density remains unchanged before and after the annihilation.  The total entropy is still conserved as axions are created with finite entropy arising from an axion chemical potential.  The chemical potential is important for axions to condense into a BEC state, a subject that we will not discuss in this paper.

For a finite change of $g_d$ factor, 
we sum up many infinitesimal changes to obtain the total converted energy density  
per $(\pi^2/30)\bar T^4$, and therefore 
\begin{equation}
rg_{d} = ({1\over 3})\Delta g_d,
\end{equation}
where $\Delta g_d = g_d - g_{d,f}$ and $g_{d,f}$ is the final value after particle annihilation. 

Now one can see that $rg_d$ has a lower bound.  The annihilated degrees of freedom $\Delta g_d$ cannot be $\ll 1$; otherwise, too little particle annihilation would have taken place, and instead of axions, the dark fermions would have become the primary dark matter.  On the other hand, $rg_d$ has an upper bound, since we have assumed the degree of freedom of dark fermions is not large.  Therefore, we have $r g_d=O(1)$, and the scales of the axion mass $m_a$ 
and the dark particle mass $\Lambda$ are confined to $O(10^{-22})$ eV and $O(100)$ eV, respectively. 

\section{Discussions}
\label{sec:4}

From Eq.(11) and given the axion mass
$m_a = 1-2 \times 10^{-22}$ eV, thus $r g_{d}\sim 0.6-0.43$, we can further constrain the model details. 
One arrives at $(8/7)\Delta g_d \sim 2.1-1.8$ for fermions.  Bosons and anti-bosons are likely ruled out since 
bosons are not known to possess a CP violating phase. 
A non-integer degrees of freedom for annihilated particles mean that the annihilation is not $100\%$ before decoupling and a fraction of particles can survive.  The annihilated degrees of freedom must be close to an integer from below for the axion to be the primary dark matter, and it suggests the dark particles have 2 degrees of freedom.

In contrast to a normal fermion 
that has 4 degrees of freedom, the only known fermions with two degrees of freedom are neutrinos.  Thus the dark fermions may be the right-handed neutrino and the left-handed anti-neutrino, despite unusually strong interactions are required between them to be annihilated. 
The non-integer degrees of freedom posit an interesting possibility for the detection of warm dark matter.
If warm dark matter is the massive sterile neutrino, it may decay to the active neutrino through emission of two photons.  Hence the two-photon line is at $\Lambda/2 \sim 45-65$ eV.   

A massless gauge boson $F^{\mu\nu}$ must exist in any viable axion model both
to couple to the axion and to account for the final heat bath.
The simplest choice is the SU(2) gauge boson that has a BRST instanton solution \cite{bel} with 3 degrees of freedom.  The dark radiation, composed of a dilute instanton gas, will assume a small fraction, $((3+\Delta g_d)/3.36)(\bar T_m/T_m)^4 \sim  2.5\%$ for $\Delta g_d\sim 2$, of the total radiation.\footnote{We have set the instanton temperature to equal $((3+\Delta g_d)/3)^{1/4}\bar T_m$, due to the fact that annihilation of dark particles leads to an increase of temperature to keep the thermal energy unchanged in the dark sector. (Also see the discussion below Eq.(10).)}   This amount may yield a detectable contribution to the cosmic microwave background radiation.  The 2013 Planck results claim non-detection of relativistic dark relics in addition to three species of neutrinos, despite that a 1-$\sigma$ excess is seen in the combined analysis \cite{pla}.  Clearly more precision data are
needed to pin down this issue.

\acknowledgements{
This work is supported by MOST of Taiwan
under the grants NSC100-2112-M-002-018-MY3.}

\newcommand {\apjl}  {Astrophys. J. Lett.}
\newcommand {\mnras} {Mon. Not. R. Astron. Soc.}
\newcommand {\na}    {New Astron.}


\begin{thebibliography}{}
\bibitem [1]{sch1} Schive, H.Y., Chiueh, T., Broadhust, T., 2014, Nature Phys., {\bf10}, 496
\bibitem [2]{sch2} Schive, H.Y., et. al, 2014, arXiv:1407- 7762
\bibitem [3]{woo} Woo, T.B., Chiueh, T., 2009, ApJ, {\bf 697}, 850
\bibitem [4]{hu} Hu, W., Barkana, R., Gruzinov, A., 2000, PRL, {\bf 85}, 1158
\bibitem [5]{lee} Lee, J.W., Lim, S., 2010, JCAP01, 007
\bibitem [6]{lor} Lora, V., Magana, J., Bernal, A., Sanchez-Salcedo, F.J., Grebel, E.K., 2012, JCAP02, 011
\bibitem [7]{har} Harko, T., 2011, JCAP05, 022
\bibitem [8]{mag} Magana, J., Matos, T., 2012, J.Phys: Conf.Ser., {\bf 378}, 012012
\bibitem [9]{rob} Robles, V.H., Matos, T., 2012, MNRAS, {\bf 422}, 282
\bibitem [10]{mar} Marsh, D.J.E., Silk, J., 2014, MNRAS, {\bf 437}, 2652
\bibitem[11]{mac} Macci`o, A.V., Paduroiu, S., Anderhalden, D., Schneider, A., Moore, B., 2012, MNRAS, {\bf 424}, 1105
\bibitem[12]{sel} Seljak, U., Makarov, A., McDonald, P., Trac, H., 2006, PRL, {\bf 97}, 1303
\bibitem [13]{vie} Viel, M., Becker, G.D., Bolton, J.S., Haehnelt, M.G., 2013, Phys.Rev.D, {\bf 88}, 043502
\bibitem [14]{asi} Arvanitaki, A., Dimopoulos, S. Dubovsky, S., Kaloper, N., March-Russell, J., 2010, PRD, {\bf 81}, 123530
\bibitem [15]{ach} Acharya, B., Bobkov, K., Kumar, P., 2010, JHEP, {\bf 1011}, 105
\bibitem[16]{pec1} Peccei, R.D., Quinn, H.R., 1977, PRL, {\bf 38}, 1440
\bibitem[17]{pec2} Peccei, R.D., Quinn, H.R., 1977, PRD, {\bf 16}, 1791
\bibitem[18]{wei} Weiberg, S., 1978, PRL, {\bf 40}, 223
\bibitem[19]{wil} Wilczek, F., 1978, PRL, {\bf 40}, 279
\bibitem[20]{kim} Kim, J.E., 1979, PRL, {\bf 43}, 103
\bibitem[21]{shi} Shifman, M., Vainshtein, A., Zakharov, V., 1980, Nucl. Phys. B{\bf 166}, 493
\bibitem[22]{din} Dine, M., Fischler, W., Srednicki, M., 1981, Phys. Lett. B{\bf 104}, 199
\bibitem[23]{zhi} Zhitnitsky, A., 1980, Sov. J. Nucl. Phys. {\bf 31},  260
\bibitem[24]{kim1} Kim, J.E., Semertzidis, Y., Tsujikawa, S., 2014, arXiv:1409-2497 
\bibitem[25]{bae} Bae, K., Huh, J., Kim, J.E., 2008, JCAP09, 005
\bibitem[26]{bel} Belavin, A.A., Polyakov, A.M., Schwartz, A.S., Tyupkin, Yu.S., 1975, Phys.Lett. B{\bf 59}, 85 
\bibitem[27]{pla} Planck Collaboration, Planck 2013 Results XVI: Cosmological Parameters, 2014, A{\&}A



\end{thebibliography}
\end{document}